\documentclass{article}

\usepackage{microtype}
\usepackage{subfigure}
\usepackage{booktabs} % for professional tables

\usepackage{hyperref}
\usepackage{graphicx}
\usepackage{xfrac}
  \PassOptionsToPackage{numbers, compress}{natbib}
\usepackage[final]{neurips_2024}

\usepackage{amsmath}
\usepackage{amssymb}
\usepackage{mathtools}
\usepackage{amsthm}
\usepackage{natbib}
\usepackage{tabularx}
\usepackage{xcolor}         % colors
\usepackage{multirow}
\usepackage{amsmath,amsfonts,bm}

\def\eqref#1{~\ref{#1}}
\def\1{\bm{1}}

\def\vtheta{{\bm{\theta}}}

\def\vf{{\bm{f}}}

\def\vx{{\bm{x}}}
\def\vy{{\bm{y}}}

\def\mPhi{{\bm{\Phi}}}

\DeclareMathAlphabet{\mathsfit}{\encodingdefault}{\sfdefault}{m}{sl}
\SetMathAlphabet{\mathsfit}{bold}{\encodingdefault}{\sfdefault}{bx}{n}

\title{Contrastive losses as generalized models of global epistasis}

\author{%
  David H. Brookes \thanks{david.brookes@dynotx.com} \\
  Dyno Therapeutics\\
  \And
  Jakub Otwinowski \\
  Dyno Therapeutics\\
  \And
  Sam Sinai \\
  Dyno Therapeutics \\
}
\date{\vspace{-5ex}}

\begin{document}

\maketitle

\begin{abstract}
   Fitness functions map large combinatorial spaces of biological sequences to properties of interest. Inferring these multimodal functions from experimental data is a central task in modern protein engineering. Global epistasis models are an effective and physically-grounded class of models for estimating fitness functions from observed data. These models assume that a sparse latent function is transformed by a monotonic nonlinearity to emit measurable fitness. Here we demonstrate that minimizing supervised contrastive loss functions, such as the Bradley-Terry loss, is a simple and flexible technique for extracting the sparse latent function implied by global epistasis. We argue by way of a fitness-epistasis uncertainty principle that the nonlinearities in global epistasis models can produce observed fitness functions that do not admit sparse representations, and thus may be inefficient to learn from observations when using a Mean Squared Error (MSE) loss (a common practice). We show that contrastive losses are able to accurately estimate a ranking function from limited data even in regimes where MSE is ineffective and validate the practical utility of this insight by demonstrating that contrastive loss functions result in consistently improved performance on benchmark tasks.
\end{abstract}

\section{Introduction}

A fitness function maps biological sequences to relevant scalar properties of the sequences, such as binding affinity to a target molecule, or fluorescent brightness. Biological sequences span combinatorial spaces and fitness functions are typically multi-peaked, due to interactions between positions in a sequence. Learning fitness functions from limited experimental data (often a minute fraction of the possible space) can be a difficult task but allows one to predict properties of sequences. These predictions can help identify promising new sequences for experimentation \citep{Wu2019} or to guide the search for optimal sequences \citep{Brookes2019, Bryant2021}. 

Even in the case of where experimental measurements are available for every possible sequence in a sequence space, inferring a model of the fitness function can be valuable for understanding the factors that impact sequences' fitness \citep{Ding2022} or how evolution might progress over the fitness landscape \citep{Wu2016}.

Numerous methods have been developed to estimate fitness functions from experimental data, including classical machine learning techniques \citep{Yang2019}, deep learning approaches \citep{Gelman2021}, and semi-supervised methods \citep{Hsu2022}. Additionally, there are many methods that incorporate biological assumptions into the modeling process, such as parameterized biophysical models \citep{Otwinowski2018biophys}, non-parametric techniques \citep{Zhou2020, Zhou2022}, and methods for spectral regularization of neural networks \citep{Aghazadeh2021}. These latter approaches largely focus on accurately modeling the influence of ``epistasis" on fitness functions, which refers to statistical or physical interactions between genetic elements, typically either amino-acids in a protein sequence or genes in a genome.

``Local" epistasis refers to interactions between a limited number of specific positions in a sequence, and is often modeled using interaction terms in a linear model of a fitness function \citep{Otwinowski2014}. ``Global" epistasis, on the other hand, refers to the presence of nonlinear relationships that affect the fitness of sequences in a nonspecific manner. A model of global epistasis typically assumes a simple latent fitness function is transformed by a monotonically increasing nonlinearity to produce observed fitness data \citep{Sailer2017, Otwinowski2018, Tareen2022, reddy2021global, Sarkisyan2016}. Typically, these models assume a particular parametric form of the latent fitness function and nonlinearity, and fit the parameters of both simultaneously. It is most common to assume that the underlying fitness function includes only additive (non-interacting) effects \citep{Otwinowski2018}, though pairwise interaction effects have been added in some models \citep{Tareen2022}. 

Despite their relative simplicity, global epistasis models have been found to be effective at modeling experimentally observed fitness functions \citep{Sarkisyan2016, Pokusaeva2019, reddy2021global}. Further, global epistasis is not just a useful modeling choice, but a physical phenomenon that can result from features of a system's dynamics \citep{Husain2020} or the environmental conditions in which a fitness function is measured \citep{Otwinowski2018}. Therefore, even if one does not use a standard global epistasis model, it is still important to consider the effects of global epistasis when modeling fitness functions.

Due to the monotonicity of the nonlinearity in global epistasis models, the latent fitness function in these models can be interpreted as a parsimonious ranking function for sequences. Herein we show that fitting a model to observed fitness data by minimizing a supervised contrastive, or ranking, loss is a simple and effective method for extracting such a ranking function. We particularly focus on the Bradley-Terry loss \citep{Bradley1952}, which has been widely used for learning-to-rank tasks \citep{Burges2005}, and more recently for ordering the latent space of a generative model for protein sequences \citep{Chan2021}. Minimizing this loss provides a technique for modeling global epistasis that requires no assumptions on the form of the nonlinearity or latent fitness functions, and can easily be applied to any set of observed fitness data. 

Further, we use an entropic uncertainty principle to show that global epistasis can result in observed fitness functions that cannot be represented using a sparse set of epistatic interactions. In particular, this uncertainty principle shows that a fitness function that is sufficiently concentrated in the fitness domain--meaning that a small number of sequences have fitness values with relatively large magnitudes--can not be concentrated in the Graph Fourier bases that represent fitness functions in terms of local epistatic interactions \citep{Stadler1995, Weinberger1991, Brookes2022}. We show that global epistasis nonlinearities tend to concentrate observed fitness functions in the fitness domain, thus preventing a sparse representation in the epistatic domain. This insight has the implication that observed fitness functions that have been affected by global epistasis may be difficult to estimate with undersampled training data and a Mean Squared Error (MSE) loss. We hypothesize that estimating the latent ranking fitness function using a contrastive loss can be done more data-efficiently than estimating the observed fitness function using MSE, and conduct simulations that support this hypothesis. Additionally, we demonstrate the practical importance of these insights by showing that models trained with the Bradley-Terry loss outperform those trained with MSE loss on nearly all FLIP benchmark tasks \citep{Dallago2021}.

% Prior to presenting our results, we first review the relevant concepts regarding fitness functions, global epistasis, and contrastive losses.

\section{Background}

\subsection{Fitness functions and the Graph Fourier transform}

A fitness function $f: \mathcal{S} \rightarrow \mathbb{R}$ maps a space of sequences $\mathcal{S}$ to a scalar property of interest. In the case where $\mathcal{S}$ contains all combinations of elements from an alphabet of size $q$ at $L$ sequence positions, then the fitness function can be represented exactly in terms of increasing orders of local epistatic interactions. For binary sequences ($q=2$), this representation takes the form:
\begin{equation*}
    f(\vx) = \beta_0 + \sum_{i=1}^L \beta_i x_i + \sum_{i j} \beta_{ij} x_i x_j + \sum_{ijk}\beta_{ijk} x_i x_j x_k + ...,
\end{equation*}
where $x_i \in \{-1, 1\}$ represent elements in the sequence and each term in the expansion represents a (local) epistatic interaction with weight $\beta_{\{i\}}$, with the expansion continuing up to $L^{\text{th}}$ order terms. Analogous representations can be constructed for sequences with any size alphabet $q$ using Graph Fourier bases\citep{Stadler1995, Weinberger1991, Brookes2022}. These representations can be compactly expressed as:
\begin{equation}\label{eq:graph_fourier}
    \vf = \mPhi \bm{\beta},
\end{equation}
where $\vf$ is a length $q^L$ vector containing the fitness values of every sequence in $\mathcal{S}$, $\mPhi$ is a $q^L \times q^L$ orthogonal matrix representing the Graph Fourier basis, and $\bm{\beta}$ is a length $q^L$ vector containing the weights corresponding to all possible epistatic interactions. We refer to $\vf$ and $\bm{\beta}$ as representing the fitness function in the fitness domain and the epistatic domain, respectively. Note that we may apply the inverse transformation of Eq.\eqref{eq:graph_fourier} to any complete observed fitness function, $\vy$ to calculate the epistatic representation of the observed data, $\bm{\beta}_{\vy} = \mPhi^T\vy$. Similarly, if $\hat{\vf}$ contains the predictions of a fitness model for every sequence in a sequence space, then $\hat{\bm{\beta}} = \mPhi^T\hat{\vf}$ is the epistatic representation of the model.

A fitness function is considered sparse, or concentrated, in the epistatic domain when $\bm{\beta}$ contains a relatively small number of elements with large magnitudes, and many elements equal to zero or with small magnitudes. In what follows, we may refer to a fitness function that is sparse in the epistatic domain as simply being a ``sparse fitness function".  A number of experimentally-determined fitness functions have been observed to be sparse in the epistatic domain \citep{Poelwijk2019, Eble2020, Brookes2022}.  Crucially, the sparsity of a fitness function in the epistatic domain determines how many measurements are required to estimate the fitness function using Compressed Sensing techniques that minimize a MSE loss function \citep{Brookes2022}. Herein we consider the effect that global epistasis has on a sparse fitness function. In particular, we argue that global epistasis results in observed fitness functions that are dense in the epistatic domain, and thus require a large amount of data to accurately estimate by minimizing a MSE loss function. However, in these cases, there may be a sparse ranking function that can be efficiently extracted by minimizing a contrastive loss function. 

\subsection{Global epistasis models}

A model of global epistasis assumes that noiseless fitness measurements are generated according to the model:
\begin{equation}\label{eq:ge_model}
    y = g\left(f(\vx)\right),
\end{equation}
where $f$ is a latent fitness function, $g$ is a monotonically increasing nonlinear function. In most cases, $f$ is assumed to include only first or second order epistatic terms and the nonlinearity is explicitly parameterized using, for example, spline functions \citep{Otwinowski2018} or sums of hyperbolic tangents \citep{Tareen2022}. The restriction that $f$ includes only low-order terms is somewhat arbitrary, as higher-order local epistatic effects have been observed in fitness data (see, e.g., \citet{Wu2016}). In general we may consider $f$ to be any fitness function that is sparse in the epistatic domain, and global epistasis then refers to the transformation of a sparse fitness function by a monotonically-increasing nonlinearity.

Global epistasis models of the form of Eq.\eqref{eq:ge_model} have proved effective at capturing the variation observed in empirical fitness data \citep{kryazhimskiy2014global, Sarkisyan2016, Otwinowski2018, Pokusaeva2019, reddy2021global, Tareen2022}, suggesting that global epistasis is a common feature of natural fitness functions. Further, it has been shown that global epistasis results from first-principles physical considerations that are common in many biological systems. In particular, \citet{Husain2020} show that global epistasis arises when the physical dynamics of a system is dominated by slow, collective modes of motion, which is often the case for protein dynamics. Aside from intrinsic/endogenous sources, the process of measuring fitness can also introduce nonlinear effects that are dependent on the experiment and not on the underlying fitness function. For example, fitness data is often left-censored, as many sequence have fitness that falls below the detection threshold of an assay. Finally, global diminishing-returns epistatic patterns have been observed widely in both single and multi-gene settings where the interactions are among genes rather than within a gene \citep{kryazhimskiy2014global, reddy2021global, bakerlee2022idiosyncratic}. 

Together, these results indicate that global epistasis is an effect that can be expected in empirically-observed fitness functions. In what follows, we argue that global epistasis is detrimental to effective modeling of fitness functions using standard techniques. In particular, global epistasis manifests itself by producing observed data that is dense in the epistatic domain. In other words, when observed fitness data is produced through Eq.\eqref{eq:ge_model} then the epistatic representation of this fitness function (calculated through application of Eq.\eqref{eq:graph_fourier}), is not sparse.
Further we argue that this effect of global epistasis makes it to difficult to model such observed data by minimizing standard MSE loss functions with a fixed amount of data. Further, we argue that fitting fitness models aimed at extracting the latent fitness function from observed data is a more efficient use of observed data that results in improved predictive performance (in the ranking sense).

While the models of global epistasis described thus far could be used for this purpose, they have the drawback that they assume a constrained form of both $g$ and $f$, which enforces inductive biases that may affect predictive performance. Here we propose a flexible alternative to modeling global epistasis that makes no assumptions on the form of $f$ or $g$. In particular, we interpret the latent fitness function $f$ as a parsimonious ranking function for sequences, and the problem of modeling global epistasis as recovering this ranking function. A natural method to achieve this goal is to fit a model of $f$ to the observed data by minimizing a contrastive, or ranking, loss function. These loss functions are designed to learn a ranking function and, as we will show, are able to recover a sparse fitness function that has been transformed by global epistasis to produce observed data. An advantage of this approach to modeling global epistasis is that the nonlinearity $g$ is modeled non-parametrically, and is free to take any form, while the latent fitness function can be modeled by any parametric model, for example, convolutional neural networks (CNNs) or fine-tuned language models, which have been found to perform well in fitness prediction tasks \citep{Dallago2021}. An accurate ranking model also enables effective optimization, as implied by the results in \citet{Chan2021}.

\subsection{Contrastive losses}

\begin{figure*}[t]
% \vskip 0.2in
\begin{center}
% \vspace{-0.4in}
\centerline{\includegraphics[width=1\textwidth]{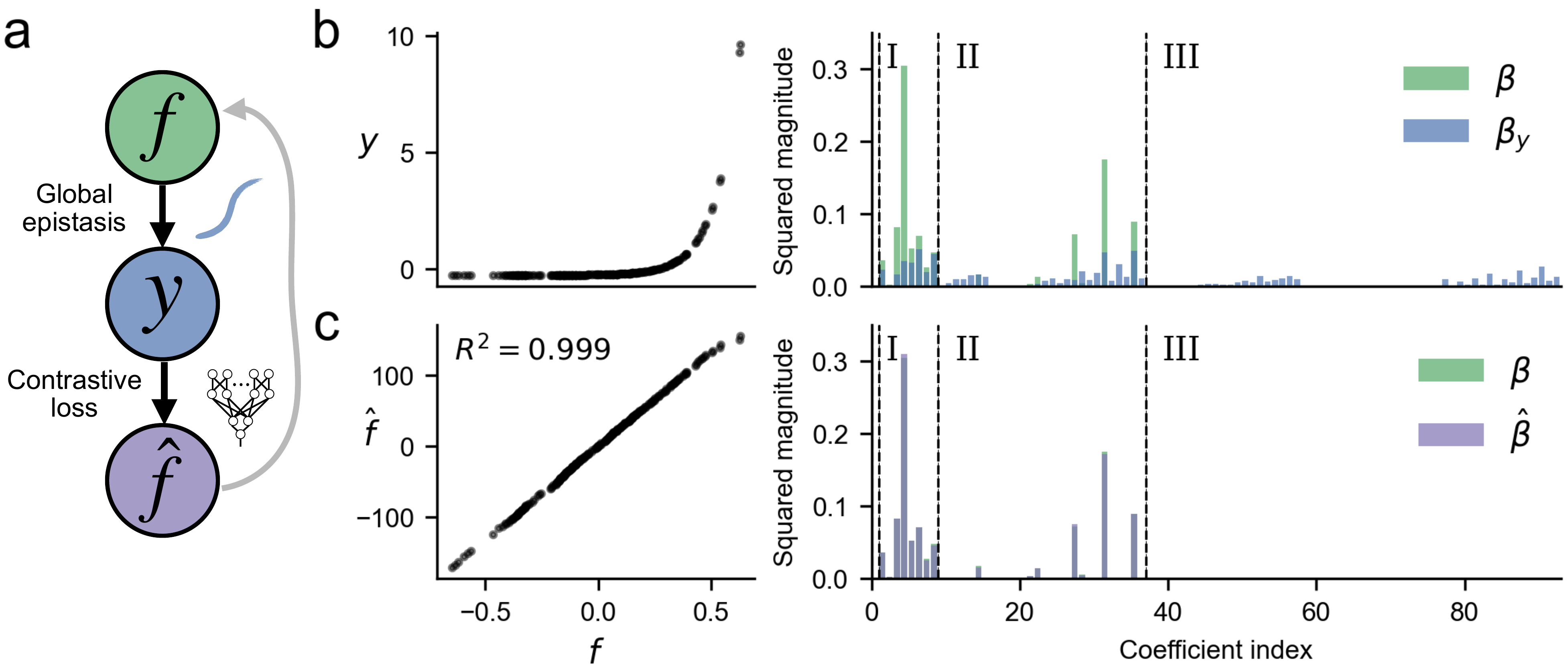}}
\vskip -0.05in
\caption{\small Recovery of latent fitness function from complete fitness data by minimizing Bradley-Terry loss. (a) Schematic of simulation. (b) Comparison between latent ($f$) and observed ($y$) fitness functions in fitness (left) and epistatic (right) domains. The latent fitness function is sampled from the NK model with $L=8$ and $K=2$ and the global epistasis function is $g(f)=\exp(10\cdot f)$. Each point in the scatter plot represents the fitness of a sequence, while each bar in the bar plot (right) represents the squared magnitude of an epistatic interaction (normalized such that all squared magnitudes sum to 1), with roman numerals indicating the order of interaction. Only epistatic interactions up to order 3 are shown. The right plot demonstrates that global epistasis produces a dense representation in the epistatic domain compared to the representation of the latent fitness in the epistatic domain. (c) Comparison between latent and estimated ($\hat{f}$) fitness functions in fitness and epistatic domains.}
\label{fig:exact_recovery}
\vskip -0.35in
\end{center}
\end{figure*}

Contrastive losses broadly refer to loss functions that compare multiple outputs of a model and encourage those outputs to be ordered according to some criteria. In our case, we desire a loss function that encourages model outputs to be ranked according to observed fitness values. An example of such a loss function is the Bradley-Terry (BT) loss \citep{Bradley1952, Burges2005}, which has the form:
\begin{equation}\label{eq:bt_loss}
    \mathcal{L}(\vtheta) = \sum_{i,j: y_i > y_j} \log\left[1+e^{-(f_{\vtheta}(\vx_i) - f_{\vtheta}(\vx_j))} \right],
\end{equation}
where $f_{\vtheta}$ is a model with parameters $\vtheta$, $\vx_i$ are model inputs and $y_i$ are the corresponding labels of those inputs. This loss compares every pair of data points and encourages the model output $f_{\vtheta}(\vx_i)$ to be greater than $f_{\vtheta}(\vx_j)$ whenever $y_i > y_j$; in other words, it encourages the model outputs to be ranked according to their labels. A number of distinct but similar loss functions have been proposed in the learning-to-rank literature \citep{Chen2009} and also for metric learning \citep{hadsell2006}. An example is the Margin ranking loss \citep{Herbrich1999}, which replaces the logistic function in the sum of Eq.\eqref{eq:bt_loss} with a hinge function.
In our experiments, we focus on the BT loss of Eq.\eqref{eq:bt_loss} as we found it typically results in superior predictive performance
% however, we also compare to models trained with the Margin loss when presenting results in benchmark tasks. 

% Both of these losses can also easily be implemented such that the sum can be quickly be calculated over all pairs of data points in a given batch.

The BT loss was recently used by \citet{Chan2021} to order the latent space of a generative model for protein sequences such that certain regions of the latent space corresponding to sequences with higher observed fitness values. In this case, the BT loss is used in conjunction with standard generative modeling losses. In contrast, here we analyze the use of the BT loss alone in order to learn a ranking function for sequences given corresponding observed fitness values.

A key feature of the contrastive loss in Eq.\eqref{eq:bt_loss} is that it only uses information about the ranking of observed labels, rather than the numerical values of the labels. Thus, the loss is unchanged when the observed values are transformed by a monotonic nonlinearity. We will show that this feature allows this loss to recover a sparse latent fitness function from observed data that has been affected by global epistasis, and enables more data-efficient learning of fitness functions compared to a MSE loss.

% \vspace{-0.1in}
\section{Results}

Our results are aimed at demonstrating three properties of contrastive losses. First, we show that given complete, noiseless fitness data (i.e. noiseless fitness values associated with every sequence in the sequence space) that has been affected by global epistasis, minimizing the BT loss enables a model to nearly exactly recover the sparse latent fitness function $f$. Next, we consider the case of incomplete data, where the aim is to predict the relative fitness of unobserved sequences. In this regime, we find through simulation that minimizing the BT loss enables models to achieve better predictive performance then minimizing the MSE loss when the observed data has been affected with global epistasis. We argue by way of a fitness-epistasis uncertainty principle that this is due to the fact that nonlinearities tend to produce fitness functions that do not admit a sparse representation in the epistatic domain, and thus require more data to learn with MSE loss. Finally, we demonstrate the practical significance of these insights by showing that minimizing the BT loss results in improved performance over MSE loss in nearly all tasks in the FLIP benchmark \citep{Dallago2021}.

% \vspace{-0.1in}
\subsection{Recovery from complete data}\label{sec:complete_recovery}

We first consider the case of ``complete" data, where fitness measurements are available for every sequence in the sequence space. The aim of our task in this case is to recover a sparse latent fitness function when the observed measurements have been transformed by an arbitrary monotonic nonlinearity. In particular, we sample a sparse fitness function $f$ from the NK model \citep{Kauffman1989}, a popular model of fitness functions that has been shown to recapitulate the sparsity properties of some empirical fitness functions \citep{Brookes2022}. The NK model has three parameters: $L$, the length of the sequences, $q$, the size of the alphabet for sequence elements, and $K$, the maximum order of (local) epistatic interactions in the fitness function. Roughly, the model randomly assigns $K-1$ interacting positions to each position in the sequence, resulting in a sparse set of interactions in the epistatic domain. The weights of each of the assigned interactions are then drawn from a independent unit normal distributions.

We then transform the sampled fitness function $f$ with a monotonic nonlinearity $g$ to produce a set of complete observed data, $y_i = g(f(\vx_i))$ for all $\vx_i \in \mathcal{S}$. The goal of the task is then to recover the function $f$ given all $(\vx_i, y_i)$  pairs. In order to do so, we model $f$ using a two layer fully connected neural network and fit the parameters of this model by performing stochastic gradient descent (SGD) on the BT loss, using the Spearman correlation between model predictions and the $y_i$ values to determine convergence of the optimization. We then compare the resulting model, $\hat{f}$, to the latent fitness function $f$ in both the fitness and epistatic domains, using the forward and inverse transformation of Eq.\eqref{eq:graph_fourier} to convert between the two domains.

Fig. \ref{fig:exact_recovery} shows the results of one of these tests. In this case, we used an exponential function to represent the global epistasis nonlinearity. The exponential function exaggerates the effects of global epistasis in the epistatic domain and thus better illustrates the usefulness of contrastive losses, although the nonlinearities in empirical fitness functions tend to have a more sigmoidal shape \citep{Otwinowski2018}. Fig. \ref{fig:exact_recovery}b shows that the global epistasis nonlinearity substantially alters the representations of the observed data $y$ in both the fitness and epistatic domains, as compared to the latent fitness function $f$. Nonetheless, Fig. \ref{fig:exact_recovery}c demonstrates that the model fitness function $\hat{f}$ created by minimizing the BT loss is able to nearly perfectly recover the sparse latent fitness function (where recovery is defined as being equivalent up to an affine transformation). This is a somewhat surprising result, as there are many fitness functions that correctly rank the fitness of sequences, and it is not immediately clear why minimizing the BT loss produces this particular sparse latent fitness function. However, this example makes clear that fitting a model by minimizing the BT loss can be an effective strategy for recovering a sparse latent fitness function from observed data that has been affected by global epistasis. Similar results from additional examples of this task using different nonlinearities and latent fitness functions are shown in Appendix \ref{sec:expanded_complete_recovery}. 

% Further, in Appendix \ref{sec:baseline_compare} we compare the results of Fig. \ref{fig:exact_recovery}c to a baseline method in which a quantile transformation is applied to the observed data.

% Appendix \ref{sec:exact_recovery_details}.

\subsection{Fitness-epistasis uncertainty principle}

\begin{figure*}[t]
% \vskip 0.2in
% \vspace{-0.4in}
\begin{center}
\centerline{\includegraphics[width=1\textwidth]{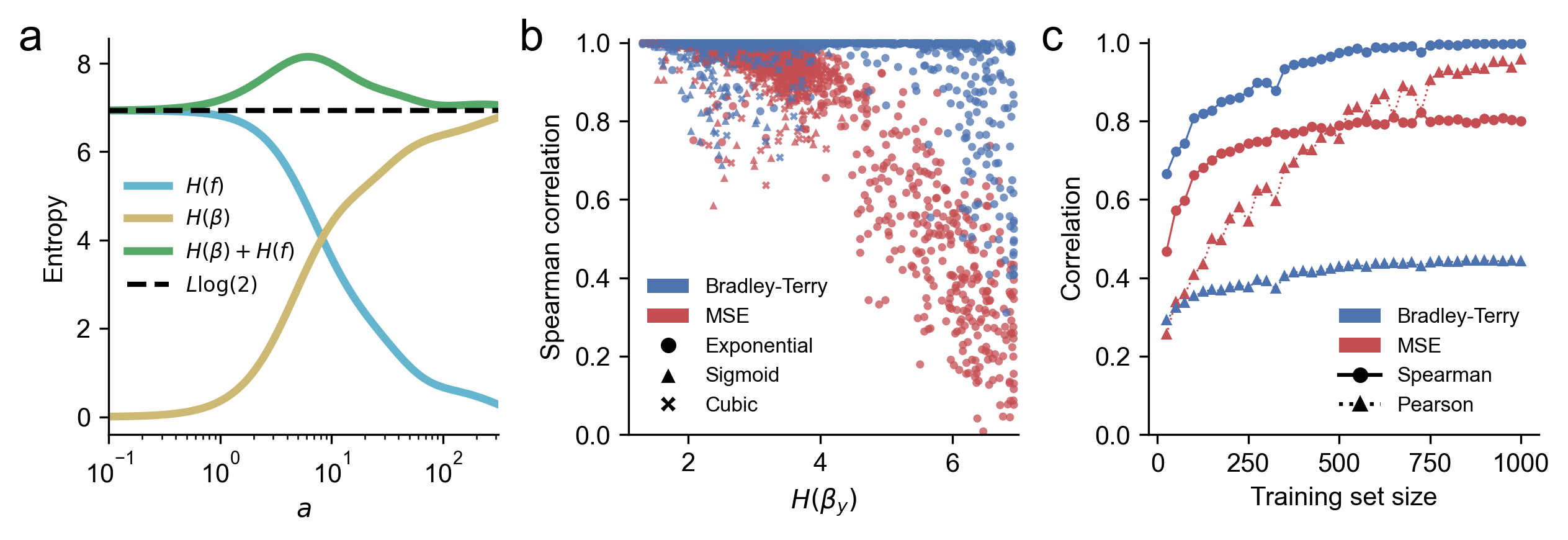}}
\vskip -0.15in
\caption{ \small (a) Demonstration of the fitness-epistasis uncertainty principle for a latent fitness function  transformed by $g(f) = \exp(a\cdot f)$ for various settings of $a$. The dashed black line indicates the lower bound on the sum of the entropies of the fitness and epistatic representations of the fitness function (b) Test-set Spearman correlation for models trained with MSE and BT losses on incomplete fitness data transformed by various nonlinearities, compared to the entropy of the fitness function in the epistatic domain. Each point corresponds to a model trained on 256 randomly sampled training points from an $L=10, K=2$ latent fitness function which was then transformed by a nonlinearity. (c) Convergence of models fit with BT and MSE losses to observed data generated by transforming an $L=10, K=2$ latent fitness function by $g(f) = \exp(10\cdot f)$. Each point represents the mean test set correlation over 200 training set replicates.}
\label{fig:incomplete_sims}
\vskip -0.35in
\end{center}
\end{figure*}

Next, we consider the case where fitness data is incomplete. Our aim is to understand how models trained with the BT loss compare to those trained with MSE loss at predicting the relative fitness of unobserved sequence using different amounts of subsampled training data. We take a signal processing perspective on this problem, and consider how the density of a fitness function in the epistatic domain affects the ability of a model to accurately estimate the fitness function given incomplete data. In particular, we demonstrate that global epistasis tends to increase the density of fitness functions in the epistatic domain, and use an analogy to Compressive Sensing (CS) to hypothesize that more data is required to effectively estimate these fitness functions when using an MSE loss \citep{Brookes2022}. In order to support this claim, we first examine the effects of global epistasis on the epistatic domain of fitness functions.

Fig. \ref{fig:exact_recovery}b provides an example where a monotonic nonlinearity applied to a sparse fitness increases the density of the the fitness function in the epistatic domain. In particular, we see that many ``spurious" local epistatic interactions must appear in order to represent the nonlinearity (e.g. interactions of order 3, when we used an NK model with $K=2$). This effect can be observed for many different shapes of nonlinearities \citep{Sailer2017, Baeza2019}. We can understand this effect more generally using uncertainty principles, which roughly show that a function cannot be concentrated on a small number of values in two different representations. In particular, we consider the discrete entropic uncertainty principle proved by \citet{Dembo1991}. When applied to the transformation in Eq.\eqref{eq:graph_fourier}, this uncertainty principle states:
\begin{equation}\label{eq: uncertainty_principle}
    H(\vf) + H(\bm{\beta}) \geq -2L \log m,
\end{equation}
where $H(\vx) = -\sum_i \frac{x^2_i}{||x||^2} \log \frac{x^2_i}{||x||^2}$ is the entropy of the normalized squared magnitudes of a vector and \mbox{$m=1 / \sqrt{q}$} when $q=2$, \mbox{$m = 1 / (q-\sqrt{q})$} when $q=3$ and \mbox{$m = 1 - 1/(q-\sqrt{q})$} otherwise. 
% The value of the right hand side of Eq.\eqref{eq: uncertainty_principle} is equal to $2 \log(\sfrac{1}{M})$, where $M$ is maximum absolute value of elements in $\Phi$, and depends only on the sequence length and size of the alphabet; in the case of binary sequences it simplifies to $L \log 2$. 
Low entropy indicates that a vector is concentrated on a small set of elements. Thus, the fitness-epistasis uncertainty principle of Eq.\eqref{eq: uncertainty_principle} shows that fitness functions cannot be concentrated in both the fitness and epistatic domains. A sparse fitness function (in the epistatic domain) must therefore be ``spread out" (i.e. dense) in the fitness domain, and vice-versa. 

The importance of this result for understanding global epistasis is that applying a nonlinearity to a dense vector will often have the effect of concentrating the vector on a smaller number of values. This can most easily be seen for convex nonlinearities like the exponential shown in Fig. \ref{fig:exact_recovery}a, but is also true of many other nonlinearities (see Theorem \ref{thm:rich_get_richer} in \ref{sec:entropy_theory} for a sufficient condition for a nonlinearity to decrease entropy in the fitness domain and \ref{sec:more_uncertainty} for additional experiments demonstrating the uncertainty principle). If this concentration in the fitness domain is sufficiently extreme, then the epistatic representation of the fitness function, $\bm{\beta}$, must be dense in order to satisfy Eq.\eqref{eq: uncertainty_principle}. Fig. \ref{fig:incomplete_sims}a demonstrates the uncertainty principle by showing how the entropy in the fitness and epistatic domains decrease and increase, respectively, as more extreme nonlinearities are applied to a sparse latent fitness function.

The uncertainty principle quantifies how global epistasis affects a fitness function by preventing a sparse representation in the epistatic domain. From a CS perspective, this has direct implications for modeling the fitness function from incomplete data. In particular, if we were to model the fitness function using CS techniques such as LASSO regression with the Graph Fourier basis as the representation, then it is well known that the number of noiseless data points required to perfectly estimate the function scales as $\mathcal{O}(S \log N)$ where $S$ is the sparsity of the signal in a chosen representation and $N$ is the total size of the signal in the representation \citep{Candes2006}. Therefore, when using these techniques, fitness functions affected by global epistasis will require more data to effectively model. Notably, the techniques for which these scaling laws apply minimize a MSE loss functions as part of the estimation procedure. Although these scaling laws only strictly apply to CS modeling techniques, we hypothesize that the intuition that fitness functions with dense epistatic representations will require more data to train an accurate model with MSE loss, even when using neural network models and SGD training procedures. In the next section we present the results of simulations that support this hypothesis by showing that the entropy of the epistatic representation is negatively correlated with the predictive performance of models trained with an MSE loss on a fixed amount of fitness data. Further, these simulations show that models trained with the BT loss are robust to the dense epistatic representations produced global epistasis, and converge faster to maximum predictive performance as they are provided more fitness data compared to models trained with an MSE loss.

% Note that there are a number of alternative uncertainty principles that could be appropriately applied instead of the entropic uncertainty principle of Eq.\eqref{eq: uncertainty_principle}, many of which are summarized by \citet{Perraudin2018}. Each of these provides the same intuition as Eq.\eqref{eq: uncertainty_principle}, but differ in how they quantify the notion of concentration of a vector. We chose to use Eq.\eqref{eq: uncertainty_principle} because entropy is perhaps the most familiar definition of concentration.

\vspace{-0.1in}
\subsection{Simulations with incomplete data}\label{sec:incomplete_data}

We next present simulation results aimed at showing that global epistasis adversely effects the ability of models to effectively learn fitness functions from incomplete data when trained with MSE loss and that models trained with BT loss are more robust to the effects of global epistasis. 

In our first set of simulations, we tested the ability models to estimate a fitness function of $L=10$ binary sequences given one quarter of the fitness measurements (256 measurements out of total of $2^{10} = 1024$ sequences in the sequence space). In each simulation, we (i) sampled a sparse latent fitness function from the NK model, (ii) produced an observed fitness function by applying one of three nonlinearities to the latent fitness function: exponential, $g(f) = \exp(a \cdot f)$, sigmoid, $g(f) =(1+e^{-a\cdot f})^{-1}$, or a cubic polynomial $g(f) = x^3 + ax$  with settings of the parameter $a$ that ensured the nonlinearity was monotonically increasing, (iii) sampled 256 sequence/fitness pairs uniformly at random from the observed fitness function to be used as training data, and (iv) trained models with this data by performing SGD on the MSE and BT losses. We ran this simulation for 50 replicates of each of 20 settings of the $a$ parameter for each of the three nonlinearities. In every case, the models were fully-connected neural networks with two hidden layers and the optimization was terminated using early stopping with $20$ percent of the training data used as a validation set. After training, we measured the extent to which the models estimated the fitness function by calculating Spearman correlation between the model predictions and true fitness values on all sequences in the sequence space. Spearman correlation is commonly used to benchmark fitness prediction methods \citep{Dallago2021, Notin2022}.

The results of each of these simulations are shown in Fig. \ref{fig:incomplete_sims}b. We see that the predictive performance of models trained with the MSE loss degrades as the entropy of the fitness function in the epistatic domain increases, regardless of the type of nonlinearity that is applied to the latent fitness function. This is in contrast to the models trained with the BT loss, which often achieve nearly perfect estimation of the fitness function even when the entropy of the fitness function in the epistatic domain approaches its maximum possible value of $L \log 2$. This demonstrates the key result that the MSE loss is sensitive to the density of the epistatic representation resulting from global epistasis (as implied by the analogy to CS), while the BT loss is robust to these effects. We additionally find that these results are maintained when the degree of epistatic interactions in the latent fitness function is changed (Appendix \ref{sec:additional_incomplete}) and when noise is added to the observed fitness functions (Appendix \ref{sec:noisy_incomplete}).

Next, we tested how training set size effects the predictive performance of models trained with MSE and BT losses on a fitness function affected by global epistasis. In order to do so, we sampled a single $L=10, K=2$ fitness function from the NK model and applied the nonlinearity $g(f)=\exp(10\cdot f)$ to produce an observed fitness function. Then, for each of a range of training set sizes between 25 and 1000, we randomly sampled a training set and fit models with MSE and BT losses using the same models and procedure as in the previous simulations. We repeated this process for 200 training set replicates of each size, and calculated both the Spearman and Pearson correlations between the resulting model predictions and true observed fitness values for all sequences in the sequence space. 

Fig. \ref{fig:incomplete_sims}c shows the mean correlation values across all 200 replicates of each training set size. There are two important takeaways from this plot. First, the BT loss achieves higher Spearman correlations than the MSE loss in all data regimes. This demonstrates the general effectiveness of this loss to estimate fitness functions affected by global epistasis. Next, we see that models trained with BT loss converge to a maximum Spearman correlation faster than models trained with MSE loss do to a maximum Pearson correlation, which demonstrates that the difference in predictive performance between models trained with MSE and BT losses is not simply due to a result of the evaluation metric being more tailored to one loss than the other. This result also reinforces our claim that fitness functions affected by global epistasis require more data to learn effectively with MSE loss, as would be predicted by CS scaling laws. 
The BT loss on the other hand, while not performant with the Pearson metric as expected by a ranking loss, seems to overcome this barrier and can be used to estimate a fitness function from a small amount of data, despite the effects of global epistasis. 
\vspace{-0.1in}
\subsection{FLIP benchmark results}\label{sec:flip_results}

% \begin{figure*}[t]
% % \vskip 0.2in
% % \vspace{-0.4in}
% \begin{center}
% \centerline{\includegraphics[width=0.87\textwidth]{figures/flip_results.png}}
% \vskip -0.15in
% \caption{ \small Comparison between MSE, Margin, and Bradley-Terry losses on FLIP benchmark tasks using the CNN baseline model. Panels correspond to tasks using the (a) GB1, (b) AAV, and (c) Thermostability datasets.  Bar height and error bars indicate the mean and standard deviation, respectively, of test set Spearman correlation over 10 random initializations of the model. Asterisks indicate statistically significant improvement over another loss function (* p<0.05, ** p<0.01, *** p<0.001, **** p<0.0001). }
% \label{fig:flip_results}
% \vskip -0.35in
% \end{center}
% \end{figure*}

% \vskip 0.15in
\begin{table*}[t]
\small
\centering
\begin{center}
% \begin{small}
\begin{sc}
\scalebox{0.86}{
\begin{tabularx}{1.15\textwidth}{Xccccc}
% \toprule
\multicolumn{2}{c}{} & \multicolumn{2}{c}{Spearman} & \multicolumn{2}{c}{Top 10\% Recall} \\
\cmidrule(lr){3-4} \cmidrule(lr){5-6} 
Data set & Split & MSE Loss & Bradley-Terry  & MSE Loss & Bradley-Terry \\
\midrule
\multirow{5}{3em}{GB1} 
& 1-vs-rest    & $0.133 \pm 0.150$ & $0.091 \pm 0.093$ & 0.097 $\pm$ 0.030 & \textbf{0.138} $\pm$ \textbf{0.051 }\\
& 2-vs-rest    & $0.564 \pm 0.026$ & $\textbf{0.607} \pm \textbf{0.009}$  & 0.250 $\pm$ 0.030 & \textbf{0.282} $\pm$ \textbf{0.008} \\
& 3-vs-rest*   & $0.814	\pm 0.049$ & $\textbf{0.880} \pm \textbf{0.003}$  & 0.539 $\pm$ 0.084 & \textbf{0.664} $\pm$ \textbf{0.014 }\\ 
& Low-vs-High* & $0.499 \pm 0.010$ & $\textbf{0.567} \pm\textbf{0.013}$ & 0.381 $\pm$ 0.028 & \textbf{0.443} $\pm$ \textbf{0.024} \\
& Sampled      & $0.930 \pm 0.002$ & $\textbf{0.951} \pm \textbf{0.002}$     & 0.823 $\pm$ 0.009 & 0.816 $\pm$ 0.010 \\
\midrule
\multirow{7}{3em}{AAV} 
& Mut-Des     & $0.751 \pm 0.006$ & $0.757 \pm 0.007$   & $0.288 \pm 0.004$ & $\textbf{0.307} \pm \textbf{0.005}$  \\
& Des-Mut     & $0.806 \pm 0.006$ & $\textbf{0.832} \pm \textbf{0.002}$  & $0.318 \pm 0.013$ & $\textbf{0.387} \pm \textbf{0.008}$ \\
& 1-vs-rest*  & $0.335 \pm 0.117$ & $\textbf{0.485} \pm \textbf{0.078}$  & $0.052 \pm 0.053$ & $\textbf{0.143} \pm \textbf{0.049}$ \\
& 2-vs-rest   & $0.748 \pm 0.010$ & $\textbf{0.798} \pm \textbf{0.003 }$   & $\textbf{0.490} \pm \textbf{0.011}$ & $0.457 \pm 0.010$ \\
& 7-vs-rest   & $0.732 \pm 0.003$ & $\textbf{0.742} \pm \textbf{0.003}$  & $0.694 \pm 0.006$ & $0.695 \pm 0.006$ \\
& Low-vs-High & $0.401 \pm 0.006$ & $\textbf{0.410} \pm \textbf{0.009} $ & $\textbf{0.180} \pm \textbf{0.009}$ & $0.170 \pm 0.006$\\
& Sampled     & $0.927 \pm 0.001$ & $\textbf{0.933} \pm \textbf{0.000}$     &  $0.650 \pm 0.005$ & $0.652 \pm 0.010$\\
\midrule
\multirow{3}{3em}{Thermostability} 
& Mixed      & $0.349 \pm 0.011$ & $\textbf{0.453} \pm \textbf{0.007}$   & $0.636 \pm 0.013$ & $0.616 \pm 0.011$ \\
& Human      & $0.511\pm 0.016$  & $\textbf{0.589} \pm \textbf{0.002}$    & $0.382 \pm	0.022$ & $\textbf{0.405} \pm \textbf{0.015}$ \\
& Human-Cell & $0.490 \pm 0.021$ & $\textbf{0.570} \pm \textbf{0.004}$ & $0.316 \pm 0.018$ & $\textbf{0.355} \pm \textbf{0.012}$ \\
\bottomrule
\end{tabularx}
}
\end{sc}
% \end{small}
\end{center}
\caption{\small Comparison between MSE and Bradley-Terry losses on FLIP benchmark tasks using the CNN baseline model. Each row represents a data set and split combination. Numerical columns indicate the mean and standard deviation of test set metrics over 10 random initializations of the model. Asterisks indicate that unmodified portions of sequences were used in training data. Bold values indicate that a loss has significantly improved performance over all other tested losses ($p < 0.05$).}
 \label{tab:flip_results_combined}
 \vspace{-0.15in}
\end{table*}

In the previous sections, we used noiseless simulated data to explore the interaction between global epistasis and loss functions. Now we present results demonstrating the practical utility of our insights by comparing the predictive performance of models trained with MSE and BT losses on experimentally-determined protein fitness data. We particularly focus on the FLIP benchmark \citep{Dallago2021}, which comprises of a total of 15 fitness prediction tasks stemming from three empirical fitness datasets. These three datasets explore multiple types of proteins, definitions of protein fitness, and experimental assays. In particular, one is a combinatorially-complete dataset that contains the binding fitness of all combinations of mutations at 4 positions to the GB1 protein \citep{Wu2016}, another contains data about the viability of Adeno-associated virus (AAV) capsids for many different sets of mutations to the wild-type capsid sequence \citep{Bryant2021}, and another contains data about the thermostability of many distantly related proteins \citep{Jarzab2020}. 

For each of the three datasets, the FLIP benchmark provides multiple train/test splits that are relevant for protein engineering scenarios. For example, in the GB1 and AAV datasets, there are training sets that contain only single and double mutations to the protein, while the associated test sets contain sequences with more than two mutations. This represents a typical situation in protein engineering where data can easily be collected for single mutations (and some double mutations) and the goal is then to design sequences that combine these mutations to produce a sequence with high fitness. In all of the FLIP tasks the evaluation metric is Spearman correlation between model predictions and fitness labels in the test set, since ranking sequences by fitness is the primary task that models are used for in data-driven protein engineering.

In the FLIP benchmark paper, the authors apply a number of different modeling strategies to these splits, including Ridge regression, training a CNN, and a number of variations on fine-tuning the ESM language models for protein sequences \citep{Rives2021}. All of these models use a MSE loss to fit the model to the data, along with any model-specific regularization losses. In our tests, we consider only the CNN model as it balances consistently high performance in the benchmark tasks with relatively straightforward training protocols, enabling fast replication with random restarts.  

We trained the CNN model on each split using the standard MSE loss and BT contrastive losses. The mean and standard deviation of Spearman correlations between the model predictions and test set labels over 10 random restarts are shown in Table \ref{tab:flip_results_combined}, third and fourth columns. By default, the FLIP datasets contain portions of sequences that are never mutated in any of the data (e.g., only 4 positions are mutated in the GB1 data, but the splits contain the full GB1 sequence of length 56). We found that including these unmodified portions of the sequence often did not improve, and sometimes hurt, the predictive performance of the CNN models while requiring significantly increased computational complexity. Therefore most of our results are reported using inputs that contain only sequence positions that are mutated in at least one train or test data point. We found that including the unmodified portions of sequences improved the performance for the Low-vs-High and 3-vs-Rest GB1 well splits, as well as the 1-vs-rest AAV split and so these results are reported in Table \ref{tab:flip_results_combined}; in these cases we found both models trained with MSE and contrastive losses had improved performance. 

Although Spearman correlation is commonly used to benchmark models of fitness functions, in practical protein engineering settings it also important to consider the ability of the model to classify the sequences with the highest fitness. To test this, we calculated the ``top 10\% recall`` for models trained with the BT and MSE losses on the FLIP benchmark data, which measures the ability of the models to correctly classify the 10\% of test sequences with the highest fitness in the test set \cite{Gelman2021}. These results are shown in the fifth and sixth columns of Table \ref{tab:flip_results_combined}. 

The results in Table\ref{tab:flip_results_combined} show that using contrastive losses (and particularly the BT loss) consistently results in improved predictive performance across a variety of practically relevant fitness prediction tasks. Further, in no case does the BT loss result in worse performance than MSE. The reasons for this result may be manifold; however, we hypothesize that it is partially a result of sparse latent fitness functions being corrupted by global epistasis Indeed, it is shown in \citet{Otwinowski2018} that a GB1 landscape closely associated with that in the FLIP benchmark is strongly affected by global epistasis. Further, many of the FLIP training sets are severely undersampled in the sense of CS scaling laws, which is the regime in which differences between MSE and contrastive losses are most apparent when global epistasis is present, as shown in Fig. \ref{fig:incomplete_sims}. In order to ensure that our results translate to additional types of fitness data outside of the FLIP benchmark, we also compared the BT and MSE losses on two of the datasets from the ProteinGym benchmark \cite{notin2023}; these corroborating results are shown in \ref{sec:proteingym_appendix}.

% \vspace{-0.1 in}
\section{Discussion}
% \vspace{-0.05in}
% Global epistasis is thought to be  a pervasive effect on experimental fitness data. Here we have shown that minimizing a contrastive loss is an effective technique for recovering a sparse latent fitness function that has been affected by global epistasis. Additionally, we have shown that global epistasis manifests itself by producing fitness data that does not admit a sparse representation in the epistatic domain. By drawing an analogy to Compressive Sensing, we then argued that a fitness function with a dense epistatic representation requires more data to effectively estimate when using an MSE loss. Our simulations show that the BT loss, however, is robust to these effects of global epistasis. The practical utility of this perspective on global epistasis and contrastive losses can be seen in our results showing that models trained with contrastive losses consistently outperform those trained with MSE loss in fitness prediction tasks on diverse set of experimentally-measured and representative protein benchmarks.

% The comparison between the Margin and BT losses in Fig. \ref{fig:flip_results} indicates that there may be cases where certain contrastive losses outperform others. This suggests that it can be fruitful in practice to compare multiple contrastive losses for any given fitness prediction task. Our primary goal is to encourage practitioners to use such contrastive losses, rather than relying on the standard MSE loss. 

Our results leave open a few avenues for future exploration. First, it is not immediately clear in what situations we can expect to observe the nearly-perfect recovery of a latent fitness function as seen in Fig. \ref{fig:exact_recovery}. A theoretical understanding of this result may either cement the promise of the BT loss, or provide motivation for the development of techniques that can be applied in different scenarios. Next, we have made a couple of logical steps in our interpretations of these results that are intuitive, but not fully supported by any theory. In particular, we have drawn an analogy to CS scaling laws to explain why neural networks trained with MSE loss struggle to learn a fitness function that has a dense representation in the epistatic domain. However, these scaling laws only strictly apply for a specific set of methods that use an orthogonal basis as the representation of the signal; there is no theoretical justification for using them to understanding the training of neural networks (although applying certain regularizations to neural network training can provide similar guarantees \citep{Aghazadeh2022}).
% there is no reason to believe that this is a privileged basis that particularly influences the training of neural networks. 
% One explanation for why neural networks struggle in the presence of global epistasis is the following: an analogous uncertainty principle to Eq.\eqref{eq: uncertainty_principle} could be constructed for any orthogonal basis that can be used to represent fitness functions; therefore a fitness function that has been sufficiently concentrated by global epistasis in the fitness domain will not admit a sparse representation in \textit{any} other representation, not just that of the Graph Fourier basis. 
Additionally, it is not clear from a theoretical perspective why the BT loss seems to be robust to the dense representations produced by global epistasis. A deeper understanding of these phenomena could be useful for developing improved techniques.

Additionally, our simulations largely do not consider how models trained with contrastive losses may be affected by the complex measurement noise commonly seen in experimental fitness assays based on sequencing counts \citep{Busia2023}. Although our simulations mostly do not consider the effects of noise (except for the simple Gaussian noise added in Appendix \ref{sec:noisy_incomplete}), our results on the FLIP benchmark demonstrate that contrastive losses can be robust to the effects of noise in practical scenarios. Further, we show in \ref{sec:noisy_toy} that the BT loss can be robust to noise in a potentially pathological scenario. A more complete analysis of the effects of noise on contrastive losses would complement these results.

% Despite these possibilities for further inquiry, our results lay the groundwork for understanding the interplay between natural properties of fitness functions and the loss functions used to train models on fitness data.

% \section*{References}

% \small
% \bibliography{main}
\bibliography{main}

\newpage
\appendix
% \onecolumn
% \section{Appendix}

\renewcommand{\thesection}{Appendix \Alph{section}}   
\setcounter{section}{0}

\renewcommand{\thesubsection}{\Alph{section}.\arabic{subsection}}

% \renewcommand{\theequation}{A\arabic{equation}}
% \setcounter{equation}{0}

% \renewcommand{\thefigure}{S\arabic{figure}}   
% \setcounter{figure}{0}

% \title{Supplementary Information: Contrastive losses as generalized models of global epistasis}
% \author{}
% \date{}
% \maketitle
% \vspace{-0.5in}
\section{Simulation details}\label{sec:sim_details}

Here we provide more specific details about the simulations used to generate Figures \ref{fig:exact_recovery} and \ref{fig:incomplete_sims} in the main text.

\subsection{Complete data simulation}\label{sec:exact_recovery_details}

This section provides more details on the simulation that is described in Section \ref{sec:complete_recovery} and Fig. \ref{fig:exact_recovery} in the main text, in which the goal was to recover a latent fitness function given complete fitness data. For this task, we first sampled a complete latent fitness function, $f(\textbf{x}_i)$ for $i=1,2,...2^L$, from the NK model, using the parameters $L=8$, $K=2$ and $q=2$. The NK model was implemented as in \cite{Brookes2022}, using the random neighborhood scheme (described in detail in Section \ref{sec:nk_def}, below) We then applied a monotonic nonlinearity to the fitness function to produce a complete set of observed data, $y_i = g(f(\textbf{x}_i))$ for $i=1,2,...2^L$. We then constructed a neural network model, $f_{\bm{\theta}}$, in which input binary sequences of length $L$ were transformed by two hidden layers with 100 nodes each and ReLU activation functions and a final linear layer that produced a single fitness output. To fit the parameters of this model, we performed stochastic gradient descent using the Adam method \cite{Kingma2014} on the Bradley-Terry (BT) loss with all $(\textbf{x}_i, y_i)$ pairs as training data, a learning rate of $0.001$, a batch size of $256$. The optimization procedure was terminated when the Spearman between the model's predictions and the observed data $y$ failed to improve over 100 epochs. Letting $\hat{\bm{\theta}}$ be the parameters of the model at the end of the optimization, we denote the estimated fitness function as $\hat{f} := f_{\hat{\bm{\theta}}}$.

In order to calculate the epistatic representations of the latent, observed and estimated fitness functions, we arranged each of these fitness functions into appropriate vectors: $\textbf{f}$, $\textbf{y}$, and $\hat{\textbf{f}}$, respectively. These vectors are arranged such that the $i^{th}$ element corresponds to the sequence represented by the $i^{th}$ row of the Graph Fourier basis. In this case of binary sequences, the Graph Fourier basis is known as the Walsh-Hadamard basis and is easily constructed, as in \cite{Aghazadeh2021} and \cite{Brookes2022}. The epistatic representations of of the latent, observed and estimated fitness functions were then calculated as $\bm{\beta} = \bm{\Phi}^T \textbf{f}$, $\bm{\beta}_y = \bm{\Phi}^T \textbf{y}$, and $\hat{\bm{\beta}} = \bm{\Phi}^T \hat{\textbf{f}}$, respectively.

\subsection{Incomplete data simulations}

This section provides further details on the simulations in Section \ref{sec:incomplete_data} and Fig. \ref{fig:incomplete_sims} where we tested the ability of models trained with MSE and BT losses to estimate fitness functions given incomplete data corrupted by global epistasis. Many of the details of these simulations are provided in the main text. We used a different set of settings of the $a$ for each nonlinearity used in the simulations shown in Fig. \ref{fig:incomplete_sims}b. In particular, for the exponential, sigmoid, and cubic functions we used 20 logarithmically spaced values of $a$ in the ranges $[0.1, 50]$, $[0.5, 25]$, and $[0.001, 10]$, respectively. In all cases where models were trained in these simulations, we used the same neural network model described in the previous section. In all of these cases, we performed SGD on either the MSE or BT loss using the Adam method with a learning rate of 0.001 and batch size equal to 64. In cases where the size of the training set was less than 64, we set the batch size to be equal to the size of the training set. In all cases, the optimization was terminated when the validation metric failed to improve after 20 epochs. The validation metrics for models trained with the BT and MSE losses were the Spearman and Pearson correlations, respectively, between the model predictions on the validation set and the corresponding labels. In order to calculate the yellow curve in Fig. \ref{fig:incomplete_sims}a and the values on the horizontal axis in Fig. \ref{fig:incomplete_sims}b, the epistatic representations of the observed fitness functions were calculated as described in the previous section. 

\subsection{Definition of the NK model}\label{sec:nk_def}

The NK model is an extensively studied random field model of fitness functions introduced by \citet{Kauffman1989}. In order to sample a fitness function from the NK model, first one chooses values of the parameters $L$, $K$, and $q$, which correspond to the sequence length, maximum degree of epistatic interaction, and size of the sequence alphabet, respectively. Next, one samples a ``neighborhood" $\mathcal{V}^{[j]}$ for each position $j$ in the sequence, which represents the $K$ positions that interact with position $j$. Concretely, for each position $j =1,...,L$, $\mathcal{V}^{[j]}$ is constructed by sampling $K$ values from the set of positions $\{1,...,L\} \setminus j$ uniformly at random, without replacement. Now let $\mathcal{S}^{(L, q)}$ be the set of all $q^L$ sequences of length $L$ and alphabet size $q$. Given the sampled neighborhoods $\{\mathcal{V}^{[j]}\}_{j=1}^L$, the NK model assigns a fitness to every sequence in $\mathcal{S}^{(L, q)}$ through the following two steps:
\begin{enumerate}
    \item Let $\textbf{s}^{[j]} \coloneqq (s_k)_{k\in\mathcal{V}^{[j]}}$
    be the subsequence of $\textbf{s}$ corresponding to the indices in the neighborhood $\mathcal{V}^{[j]}$.  Assign a `subsequence fitness', $f_j(\textbf{s}^{[j]})$ to every possible subsequence, $\textbf{s}^{[j]}$, by drawing a value from the normal distribution with mean equal to zero and variance equal to $\sfrac{1}{L}$. In other words, $f_j(\textbf{s}^{[j]})\sim \mathcal{N}(0, \sfrac{1}{L})$ for every $\textbf{s}^{[j]}\in\mathcal{S}^{(K, q)}$, and for every $j=1,2,...,L$.
    \item For every $\textbf{s}\in \mathcal{S}^{(L, q)}$, the subsequence fitness values are summed to produce the total fitness values \mbox{$f(\textbf{s}) = \sum_{j=1}^L f_j(\textbf{s}^{[j]})$}.
\end{enumerate}

% \newpage
% \clearpage
\section{Additional complete data results}

Here we provide additional results to complement those described in Section \ref{sec:complete_recovery} and shown in Fig. \ref{fig:exact_recovery}.

\subsection{Additional nonlinearities and latent fitness function}\label{sec:expanded_complete_recovery}

\begin{figure}[h]
\vskip -0.15in
\begin{center}
\centerline{\includegraphics[width=0.7\textwidth]{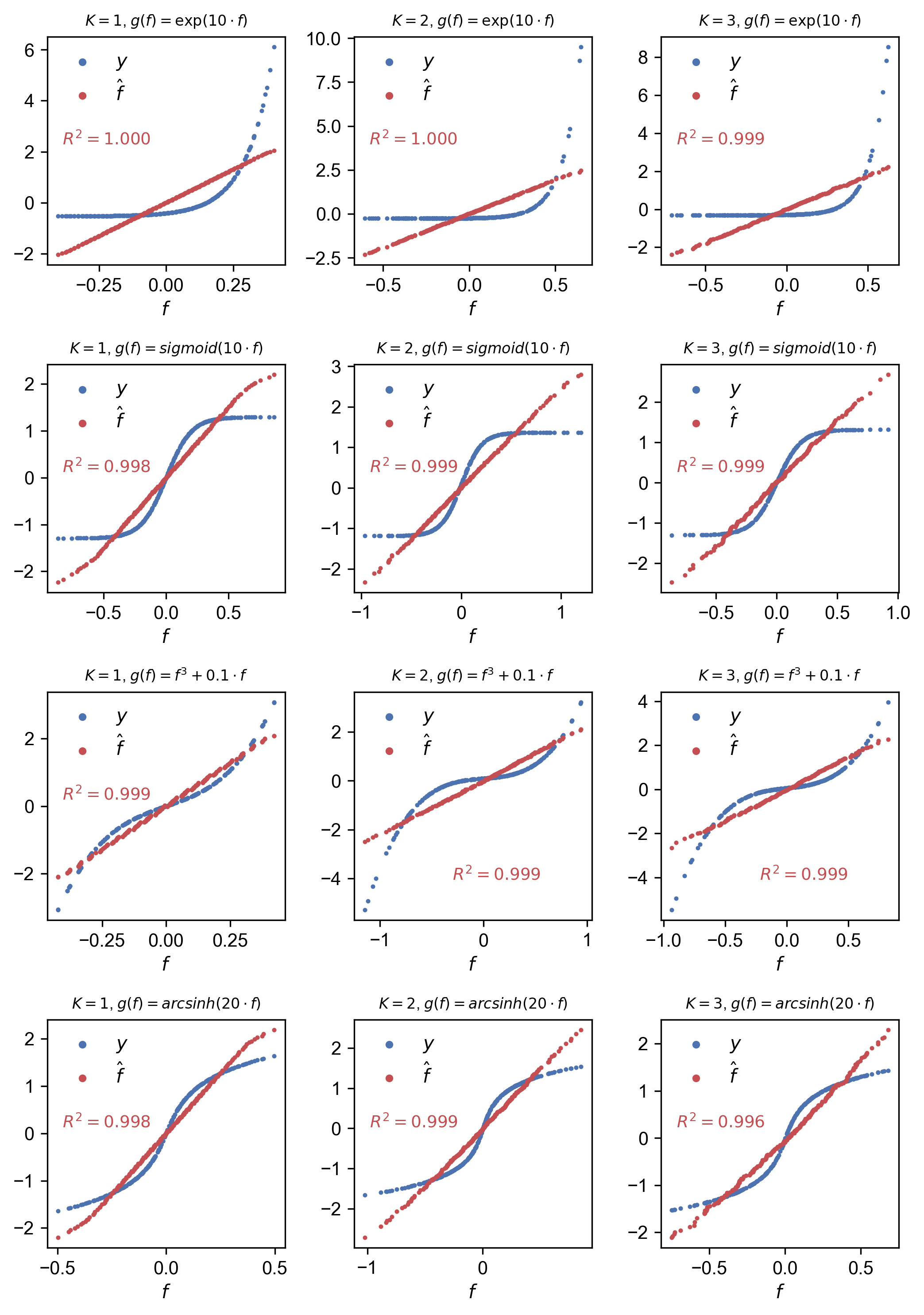}}
\vskip -0.15in
\caption{Results from multiple examples of the task of recovering a latent fitness function given complete observed data transformed by a global epistasis nonlinearity. Each sub-plot shows the results of one such task. The setting of $K$ used to sample the latent fitness function from the NK model and the particular form of the nonlinearity $g(f)$ used are indicated in each sub-plot title. The horizontal axis in each sub-plot represents the values of the latent fitness function, while the vertical axis represents the values of either the observed data (blue dots) or model predictions (red dots). For ease of plotting, all fitness functions were normalized to have an empirical mean and std. dev. of 1, respectively. The $R^2$ correlation between the latent fitness function and the model predictions are indicated in red text.
}
\label{fig:multiple_exact_recovery}
\end{center}
\end{figure}

In order to more fully demonstrate the ability of models trained with the BT loss to recover sparse latent fitness functions, we repeated the test shown in Fig. \ref{fig:exact_recovery} and described in Sections \ref{sec:complete_recovery} and \ref{sec:exact_recovery_details} for multiple examples of nonlinearities and different settings of the $K$ parameter in fitness functions sampled from the NK model. In each example, a latent fitness function of $L=8$ binary sequences was sampled from the NK model with a chosen setting of $K$ and a nonlinearity $g(f)$ was applied to the latent fitness function to produce observed data $y$. The results of these simulations are shown in Fig. \ref{fig:multiple_exact_recovery}, below. In all cases, the model almost perfectly recovers the sparse latent fitness function given complete fitness data corrupted by global epistasis.

\clearpage 
% \subsection{Comparison to quantile transformation}\label{sec:baseline_compare}

% In order to provide a baseline for the results shown in Figure \ref{fig:exact_recovery}c, we attempted to recover the latent fitness function using quantile transformations rather than using the Bradley-Terry loss. In particular, we applied a quantile transformation the observed fitness data $\textbf{y}$ to produce a vector $\textbf{f}_q$ that approximately follows either a uniform or Gaussian distribution. In both cases, we used 10 quantiles for discretization. The results of these baseline tests are shown in Figure \ref{fig:fig1_baseline}, below. By comparing to the results in Figure \ref{fig:exact_recovery}c, we can see that the quantile transformations are not as effective at recovering the latent fitness function as the Bradley-Terry loss.

% \begin{figure}[h]
% \vskip -0.15in
% \begin{center}
% \centerline{\includegraphics[width=1\textwidth]{figures/fig1_baseline.png}}
% \vskip -0.15in
% \caption{Complete data recovery from a quantile transformation to a (a) uniform distribution (b) Gaussian distribution. Plot descriptions are as in Figure \ref{fig:exact_recovery}c.
% }
% \label{fig:fig1_baseline}
% \end{center}
% \end{figure}

% \newpage
\clearpage

\section{Characterizing nonlinearities that decrease entropy}\label{sec:entropy_theory}

\newtheorem{theorem}{Theorem}
\newtheorem{lemma}{Lemma}
\newtheorem{corollary}{Corollary}
\theoremstyle{remark}

Here we provide a sufficient condition on a nonlinearity for the nonlinearity to reduce the entropy of a fitness function in the fitness domain. In other words, we characterize a class of $g$ such that $H(g(\textbf{f})) \leq H(\textbf{f})$. First, we define probability vectors $\textbf{p}$ and $\textbf{q}$ such that  $p_i \coloneqq \dfrac{f_i^2}{||\textbf{f}||^2}$ and $q_i \coloneqq \dfrac{g(f_i)^2}{||g(\textbf{f})||^2}$ for all $i=1,...,N$. 

We additionally define the vector $\textbf{w}$ such that $w_i \coloneqq \dfrac{q_i}{p_i}$ and the probability vector $\textbf{r}(\alpha)$ where $r_i(\alpha) = \dfrac{p_i w_i^{\alpha}}{\sum_{j=1}^N p_i w_i^{\alpha}}$ for all $i=1,...,N$. Note that $\textbf{r}(0) = \textbf{p}$ and $\textbf{r}(1) = \textbf{q}$.

Finally, let $H_s(\textbf{p}) = -\sum_{i=1}^N p_i \log p_i $ be the Shannon entropy of a probability vector \textbf{p}.

A note on notation: below we are primarily concerned with distributions over discrete outcomes indexed by integers. Given a probability vector $\textbf{p}$ representing such a distribution and a vector $\textbf{x}$ representing values associated with each outcome, we denote the expectation of values of $\textbf{x}$ over the distribution represented by $\textbf{p}$ as $E_p[x] = \sum_{i} p_i x_i$. Similarly, we write covariances as $\textnormal{Cov}_p(x, y) = E_p[xy] - E_p[x]E_p[y]$.

\begin{theorem}\label{thm:rich_get_richer}
Assume without loss of generality that \textbf{p} is sorted such that $p_i \geq p_{i+1}$ for \mbox{$i=1,...,N-1$}. Additionally assume that $q_i = 0$ if $p_i=0$ for $i$. If $w_i \geq w_{i+1}$ for all \mbox{$i=1,...,N-1$}, then $H(g(\textbf{f})) \leq H(\textbf{f})$.
\end{theorem}

To prove Theorem \ref{thm:rich_get_richer}, we first provide a number of lemmas.

\begin{lemma}\label{lem:shannon_deriv} The derivative of the Shannon entropy of $\textbf{r}(\alpha)$ with respect to $\alpha$ is equal to the negative covariance between $\log \textbf{r}(\alpha)$ and $\log \textbf{w}$. Specifically, the derivative is given by
\begin{equation*}
     \dfrac{\mathrm{d} H_s(\textbf{r}(\alpha))}{\mathrm{d} \alpha} = -\textnormal{Cov}_{r(\alpha)}(\log r(\alpha) , \log w).
\end{equation*}
\end{lemma}

\begin{lemma}\label{lem:cov_rearrange}
    Given a probability vector $\textbf{p}$ of length $N$ and any two vectors $\textbf{x}$ and $\textbf{y}$ of length $N$, we have
\begin{equation} \label{eq:cov_rearrange}
\textnormal{Cov}_p(x, y) = \sum_{j > i} p_i p_j (x_i - x_j)(y_i - y_j).
\end{equation}
\end{lemma}

\begin{corollary}\label{corr:cov_is_positive}
    If two vectors $x$ and $y$ of length $N$ are sorted such that $x_i \geq x_{i+1}$ and $y_i \geq y_{i+1}$ for $i=1,...,N-1$, then $\textnormal{Cov}_{p}(x, y) \geq 0$ for any probability vector $\textbf{p}$.
\end{corollary}

\begin{proof}[Proof of Theorem 1]
First we note that by the definition of entropy given in the main text, we have  $H(\textbf{f}) = H_s(\textbf{p})$ and $H(g(\textbf{f})) = H_s(\textbf{q})$. Therefore we need only prove that $H_s(\textbf{q}) \leq H_s(\textbf{p})$.

By Lemma \ref{lem:shannon_deriv}, we have that
\begin{equation}\label{eq:cov_integral}
    H_s(\textbf{q}) - H_s(\textbf{p}) = -\int_0^1 \textnormal{Cov}_{r(\alpha)}(\log r(\alpha), \log w) \mathrm{d}\alpha.
\end{equation}

Both $\textbf{p}$ and $\textbf{w}$ are sorted by assumption and therefore it is clearly true that $\textbf{r}(\alpha)$ is also sorted such that $r_i(\alpha) \geq r_{i+1}(\alpha)$ for $i=1,...,N-1$ and all $\alpha \in [0, 1]$. Since the logarithm is a monotonic function, both $\log \textbf{r}(\alpha)$ and $\log \textbf{w}$ are similarly sorted. Therefore, by Corollary \ref{corr:cov_is_positive}, we have that the integrand in Eq. \eqref{eq:cov_integral} is positive for all $\alpha \in [0,1]$. Therefore, the integral must be positive and $H_s(\textbf{q}) - H_s(\textbf{p}) < 0$.

\end{proof}

\begin{proof}[Proof of Lemma \ref{lem:shannon_deriv}]
\begin{align*}
    \dfrac{\mathrm{d} H_s(\textbf{r}(\alpha))}{\mathrm{d} \alpha} & = - \sum_{i=1}^N (1+\log r_i(\alpha))\dfrac{\mathrm{d} r_i(\alpha)}{\mathrm{d} \alpha} \\ 
    &= -\sum_{i=1}^N \left(1 + \log r_i(\alpha)\right) \dfrac{p_i w_i^{\alpha}}{\sum_{j=1}^N p_i w_i^{\alpha}} \left( \log w_i - \dfrac{\sum_{k=1}^N p_k w_k^{\alpha}\log w_k}{\sum_{j=1}^N p_i w_i^{\alpha}}\right) \\
    &= -\sum_{i=1}^N r_i(\alpha)(1+\log r_i(\alpha))\left(\log w_i - \sum_k r_k(\alpha) \log w_k \right) \\ 
    &= -\sum_{i=1}^{N} r_i(\alpha) \left(\log w_i + \log r_i(\alpha) \log w_i - E_{r(\alpha)}[\log w]- \log r_i(\alpha) E_{r(\alpha)}[\log w]\right) \\
    &= -E_{r(\alpha)}[\log w] - E_{r(\alpha)}[ \log r(\alpha) \log w]  + E_{r(\alpha)}[\log w]  + E_{r(\alpha)}[ \log r(\alpha)] E_{r(\alpha)}[\log w] \\
    &= - E_{r(\alpha)}[ \log r(\alpha) \log w] + E_{r(\alpha)}[ \log r(\alpha)] E_{r(\alpha)}[\log w] \\
    &= -\text{Cov}_{r(\alpha)}(\log r(\alpha) , \log w)
\end{align*}
\end{proof}

\begin{proof}[Proof of Lemma \ref{lem:cov_rearrange}]
\begin{align*}
    &\sum_{i=1}^N \sum_{j=i}^N p_i p_j (x_i - x_j)(y_i - y_j) =\dfrac{1}{2}\sum_{i=1}^N \sum_{j=1}^N p_i p_j (x_i - x_j)(y_i - y_j) \\
    &= \dfrac{1}{2}\sum_{i=1}^N p_i x_i y_i \sum_{j=1}^N p_j - \dfrac{1}{2}\sum_{i=1}^N p_i x_i \sum_{j=1}^N p_j y_j - \dfrac{1}{2}\sum_{i=1}^N p_i y_i \sum_{j=1}^N p_j x_j + \dfrac{1}{2} \sum_{j=1}^N p_j x_j y_j \sum_{i=1}^N p_i \\
    &= E_p[xy] - E_p[x]E_p[y] \\
    &= \textnormal{Cov}_p(x, y)
\end{align*}
where the first line follows from the summand equaling zero when $i=j$.
\end{proof}

\begin{proof}[Proof of Corollary \ref{corr:cov_is_positive}] This is a straightforward consequence of Lemma \ref{lem:cov_rearrange}. For every term in the sum of Eq. \eqref{eq:cov_rearrange}, we have that $j>i$, and therefore $x_i - x_j \geq 0$ and $y_i - y_j \geq 0$ by assumption. Further $\textbf{p}$ is a probability vector and thus $p_i > 0$ for all $i$. Therefore every element of the sum is positive and the covariance is positive.

\end{proof}

Notably, the condition of Theorem \ref{thm:rich_get_richer} is not necessary in order decrease entropy in the fitness domain. Indeed, many of the nonlinearities tested herein do not satisfy this condition. Eq. \eqref{eq:cov_integral} is an illuminating relationship that may provide further insight into additional classes of nonlinearities that decrease entropy.

\clearpage

\section{Additional examples of uncertainty principle}\label{sec:more_uncertainty}

Here we provide additional examples showing that nonlinearities tend to decrease the entropy in the fitness domain of sparse fitness, which causes corresponding increases in the entropy in the epistatic domain due to the uncertainty principle of Eq. \eqref{eq: uncertainty_principle}. We show this for four nonlinearities:exponential, $g(f) = \exp(a \cdot f)$, sigmoid, $g(f) =(1+e^{-a\cdot f})^{-1}$, cubic polynomial $g(f) = x^3 + ax$ with $a>0$, and a hinge function that represents the left-censoring of data, $g(f) = \max(0, f-a)$. Left-censoring is common in fitness data, as biological sequences will often have fitness that falls below the detection threshold of an assay. For each of these nonlinearities, we chose a range of settings of the $a$ parameter, and for each setting of the $a$ parameter, we sampled 200 fitness functions from the NK model with parameters $L=8, K=2$ and $q=2$ and transformed each function by the nonlinearity with the chosen setting of $a$. For each of these transformed fitness functions, we calculated the entropy of the function in the fitness and epistatic domains. The mean and standard deviation of these entropies across the fitness function replicates are shown in Fig. \ref{fig:expanded_uncertainty}, below. In each case, we can see that the nonlinearities cause the fitness domain to become increasingly concentrated, and the epistatic domain to become increasingly sparse.

\begin{figure}[h]
% \vskip 0.2in
\begin{center}
\centerline{\includegraphics[width=0.6 \textwidth]{figures/expanded_uncertainty.png}}
\caption{Demonstration of the fitness-epistasis uncertainty principle for multiple examples of nonlinearities. The title of the subplot indicates the nonlinearity used to produce the results in that subplot. The lines and shaded regions represent the mean and std. dev. of entropies, respectively, across 200 replicates of latent fitness functions sampled from the NK model. The black dotted line indicates the lower bound on the sum of the entropies in Eq. \eqref{eq: uncertainty_principle}.
}
\label{fig:expanded_uncertainty}
% \vskip -0.3in
\end{center}
\end{figure}

\clearpage
\section{Additional incomplete data results}\label{sec:incomplete_appendix}

Here we provide additional results to complement those described in Section \ref{sec:incomplete_data} and shown in Fig. \ref{fig:incomplete_sims}.

\subsection{Additional nonlinearity and latent fitness function parameters}\label{sec:additional_incomplete}

\begin{figure*}[h!]
% \vskip 0.2in
% \vspace{-0.4in}
\begin{center}
\centerline{\includegraphics[width=0.75\textwidth]{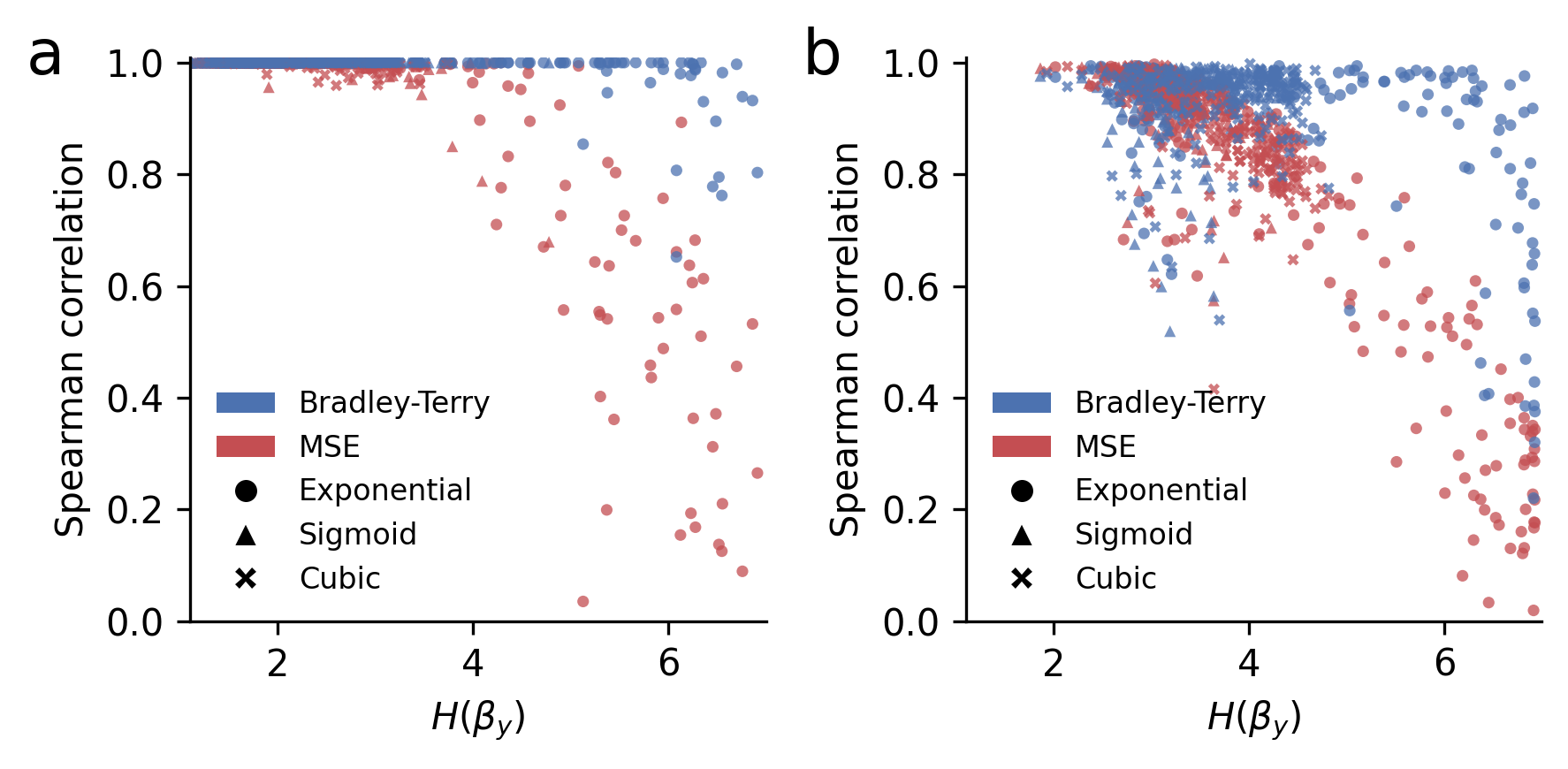}}
\vskip -0.15in
\caption{ \small Results from incomplete data simulations for latent fitness functions drawn from the NK model with (a) $K=1$ and (b) $K=2$. Plot descriptions are as in Figure \ref{fig:incomplete_sims}b. }
\label{fig:additional_entropy_sims}
\vskip -0.35in
\end{center}
\end{figure*}

Here we show results for incomplete data simulations using latent fitness functions with varying orders of epistatic interactions and for additional parameterizations of the global epistasis nonlinearity. In particular, we repeated the simulations whose results are shown in Fig. \ref{fig:incomplete_sims}b using latent fitness functions drawn from the NK model with $K=1$ and $K=3$, with all other parameters of the simulation identical to those used in the $K=2$ simulations described in the main text. We ran the simulations for 10 replicates of each of the 20 settings of the $a$ parameter for each of the three nonlinearities. The results of these simulations are shown in Figure \ref{fig:additional_entropy_sims}, below. These results are qualitatively similar to those in Figure \ref{fig:additional_entropy_sims}, demonstrating that the simulation results are robust to the degree of epistatic interactions in the latent fitness function.

\begin{figure*}[h]
% \vskip 0.2in
% \vspace{-0.4in}
\begin{center}
\centerline{\includegraphics[width=0.8\textwidth]{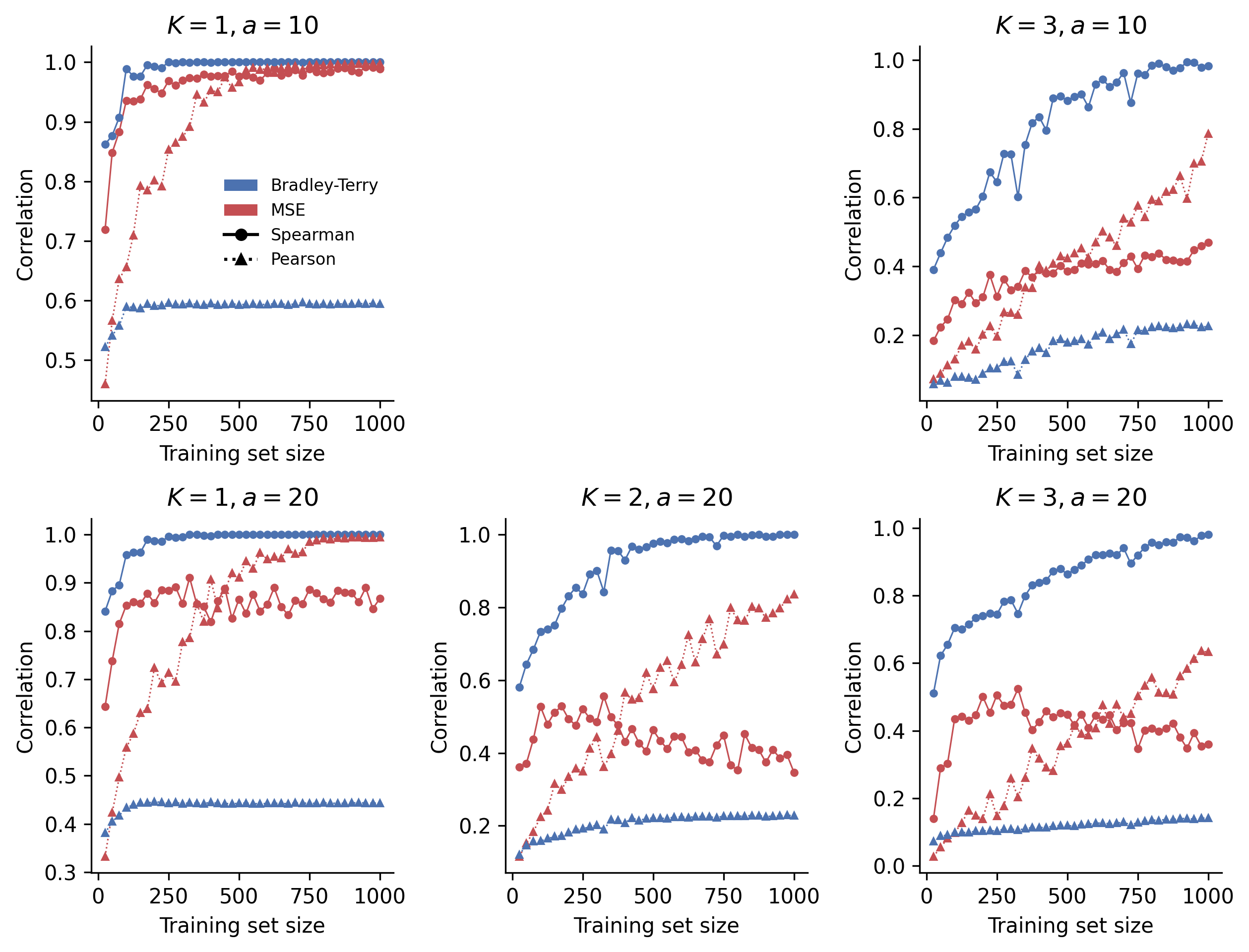}}
\vskip -0.15in
\caption{ \small Results from incomplete data simulations using multiple settings of the $K$ parameter in the NK model and the $a$ parameter in the global epistasis nonlinearity. The settings of $K$ and $a$ used to generate each subplot are shown in the title of the subplots; results are not shown for the $K=2$ and $a=10$ case because these results are shown in Figure \ref{fig:incomplete_sims}c of the main text. Plot descriptions are as in Figure \ref{fig:incomplete_sims}c; each point in the plots represents the mean over 40 replicate simulations. }
\label{fig:additional_train_size_sims}
% \vskip -0.35in
\end{center}
\end{figure*}

We also repeated the simulations whose results are shown in Figure \ref{fig:incomplete_sims}c using different settings of the $K$ parameter in the NK model, as well as multiple settings of the $a$ parameter that determines the strength of the global epistasis nonlinearity. All other parameters of the simulations are identical to those used in the $K=2$ and $a=10$ simulation described in the main text. The results of these simulations, averaged over 40 replicates for each training set size, are shown in Fig \ref{fig:additional_train_size_sims}, below. These results are qualitatively similar to those shown in Figure \ref{fig:incomplete_sims}c, demonstrating the robustness of the simulation results to different forms of latent fitness functions and nonlinearities.

\clearpage

\subsection{Adding noise to the observed data}\label{sec:noisy_incomplete}

\begin{figure*}[h]
\begin{center}
\centerline{\includegraphics[width=0.75\textwidth]{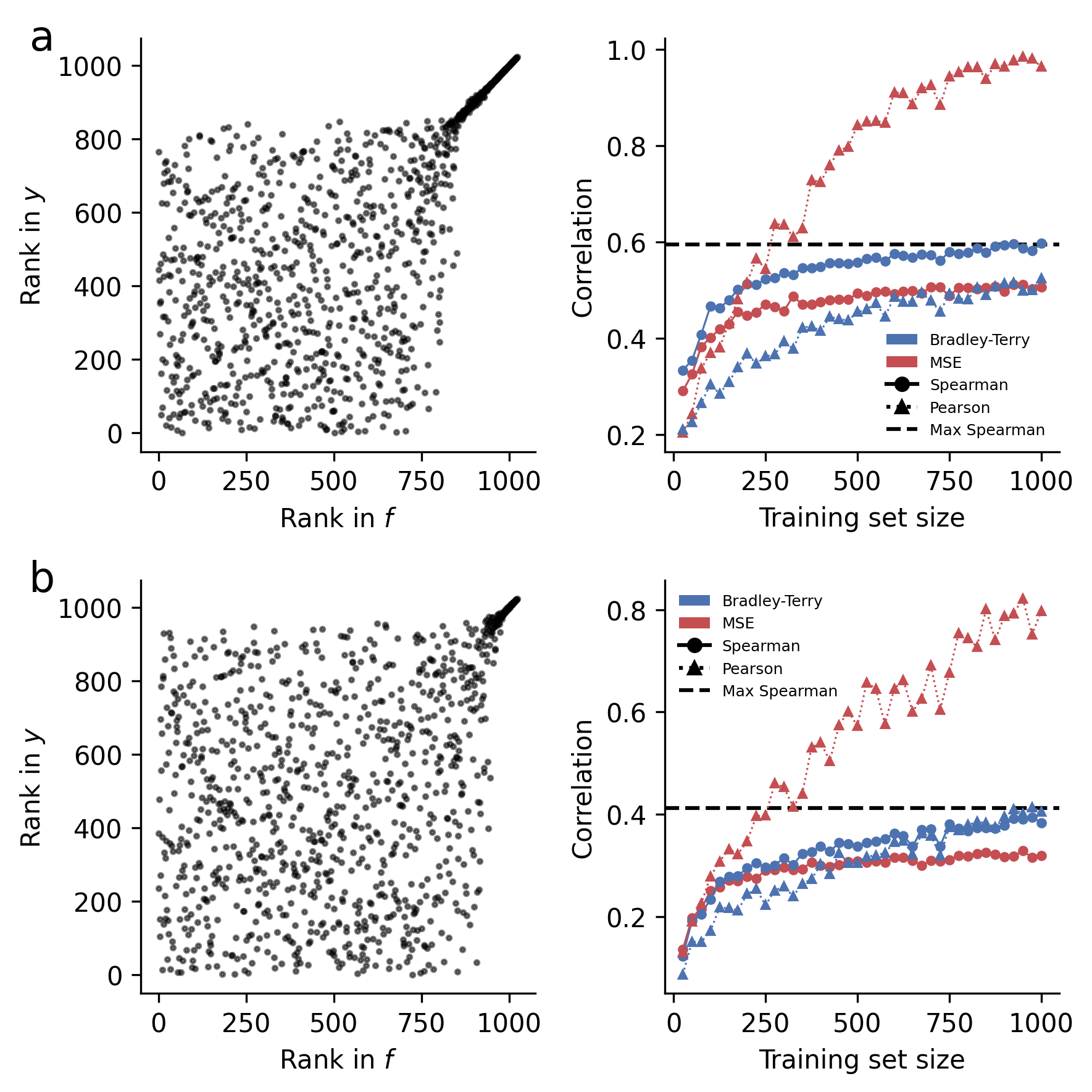}}
\vskip -0.15in
\caption{\small Incomplete data simulations with additive Gaussian noise with standard deviation equal to (a) 0.025 and (b) 0.1. Left plots show the rankings of sequences in the latent NK fitness function, $\textbf{f}$, against the ranking of sequences in the noisy observed fitness function, $\textbf{y}$. The right plot descriptions are as in Figure \ref{fig:incomplete_sims}c, with the addition of black dashed lines that indicates the Spearman correlation between the latent fitness function, $\textbf{f}$, and the noisy observed fitness function, $\textbf{y}$, which is the maximum Spearman correlation that can be achieved in this test.}
\label{fig:noisy_incomplete}
\end{center}
\end{figure*}

Here we test how the addition of homoskedastic Gaussian noise affects the results incomplete data simulations. In particular, we repeated the set of simulations in which we test the ability of the MSE and BT losses to estimate a fitness function at varying sizes of training set, but now added Gaussian noise with standard deviation of either $0.025$ or $0.1$ to the observed fitness function (first normalized to have mean and variance equal to 0 and 1, respectively). The results of these tests, averaged over 40 replicates for each training set size, are shown in Figure \ref{fig:noisy_incomplete}, below. We can see that despite the noise drastically altering the observed ranking, the BT loss still outperforms the MSE loss in terms of Spearman correlation at both noise settings. 

\clearpage

\clearpage
\section{Noisy toy simulation}\label{sec:noisy_toy}

A potential disadvantage of the BT and Margin losses used in the main text is that neither takes into consideration the size of the true observed gap between the fitness of pairs of sequences. In particular, Eq. \eqref{eq:bt_loss} shows that any two sequences in which $y_i > y_j$ will be weighted equally in the loss, regardless of the magnitude of the difference $y_i - y_j$. This has implications for how measurement noise may affect these losses. In particular, the relative ranking between pairs of sequences with a small gap in fitness in more likely to have been swapped due to measurement noise, compared to a pair of sequences with a large gap. This suggests that contrastive losses may exhibit pathologies in the presence of certain types of noise. 

Here we examine a toy scenario in which the observed fitness data has the form $y_i = I(\textbf{x}_i) + \epsilon$, where $I(\textbf{x}_i)$ is an indicator function that is equal to zero or one and $\epsilon\sim \mathcal{N}(0, \sigma^2)$ is Gaussian noise.  One may expect contrastive loses to exhibit pathologies when trained on this data because the relative rankings between sequences that have the same value of $I(\textbf{x})$ is due only to noise. 

\begin{figure}[h]
% \vskip -0.15in
\begin{center}
\centerline{\includegraphics[width=0.95\textwidth]{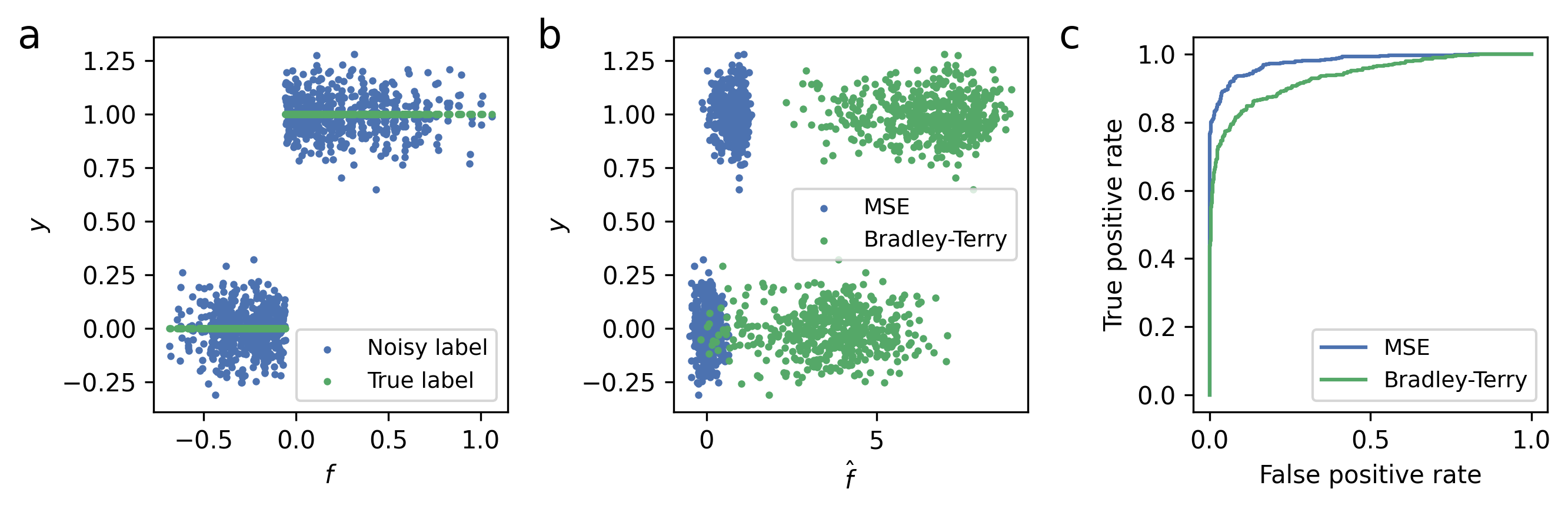}}
\vskip -0.15in
\caption{Toy simulation testing the effects of noise on the BT loss. (a) Noisy and true labels of training and test data (vertical axis), plotted against NK fitness (horizontal axis). True labels correspond to $I(\textbf{x})$ values, while noisy labels correspond to $y=I(\textbf{x})+\epsilon$ values. (b) Observed noisy labels (vertical axis) plotted against model predictions from a models trained with the MSE and BT losses (horizontal axis). (c) ROC curves comparing the ability of the models trained with MSE and BT losses to classify test sequences according to true labels. The blue (MSE) and green (BT) curves have AUC values of 0.977 and 0.932, respectively.
}
\label{fig:binary_noise_sim}
\end{center}
\end{figure}

In order to construct this scenario, we sampled an $L=10$, $K=2$ binary fitness function $f(\textbf{x}$ from the NK model, and then let $I(\textbf{x}) = 0$ when $f(\textbf{x}) < \text{med}(f)$ and 1 otherwise, where $\text{med}(f)$ is the median NK fitness of all sequences. We then added Gaussian noise with $\sigma=0.1$ to produce the observed noisy data $y$. We consider $I(\textbf{x})$ to be the true binary label of a sequence and $y$ to be the noisy label of the sequence. The relationship between the NK fitness, true labels and noisy labels is shown in Fig. \ref{fig:binary_noise_sim}a, below. Next, we randomly split the data into training and test sets containing 100 and 924 sequences and noisy labels, respectively. We used the training set to fit two neural network models, one using the MSE loss and the other using the BT loss. The models and training procedure were the same as described in \ref{sec:sim_details}. The predictions of these models on test sequences, compared to noisy labels, are shown in \ref{fig:binary_noise_sim}b. We next constructed Receiver Operating Characteristic (ROC) curves that test the ability of each model to classify sequences according to their true labels $I(\textbf{x})$ (Fig. \ref{fig:binary_noise_sim}c). These curves show that while the MSE loss does slightly outperform the BT loss in this classification task, the BT loss still results in an effective classifier and does not exhibit a major pathology in this scenario.  

Notably, on empirical protein landscapes, as shown in Table \ref{tab:flip_results_combined}, the performance gain by contrastive losses out-weighs the potential drawback of further sensitivity to this kind of noise, and contrastive losses ultimately result in improved predictive performance.

\clearpage
\section{Results on ProteinGym benchmark datasets}\label{sec:proteingym_appendix}

We compared the MSE vs Bradley-Terry loss on the \texttt{CAPSD\_AAV2S\_Sinai\_2021} and \texttt{GFP\_AEQVI\_Sarkisyan\_2016} datasets from the ProteinGym benchmark \cite{notin2023} using the same protocol as described in \ref{sec:flip_results}. These are two of the most relevant datasets in ProteinGym because they contain a large number of multi-mutants, whereas many other datasets contain only single and double mutations. The results for 10 replicates of uniform 80/20 train test splits are shown below:

\begin{table*}[h]
\small
\centering
\begin{center}
% \begin{small}
\begin{sc}
\scalebox{0.9}{
\begin{tabularx}{1.1\textwidth}{Xccccc}
% \toprule
\multicolumn{1}{c}{} & \multicolumn{2}{c}{Spearman} & \multicolumn{2}{c}{Top 10\% Recall} \\
\cmidrule(lr){2-3} \cmidrule(lr){4-5} 
Dataset & MSE Loss & Bradley-Terry  & MSE Loss & Bradley-Terry \\
\midrule
CAPSD\_AAV2S\_Sinai\_2021   & $0.912 \pm 0.003$ & $\textbf{0.920} \pm \textbf{0.003}$ & 0.915 $\pm$ 0.001 & 0.915 $\pm$ 0.001\\
GFP\_AEQVI\_Sarkisyan\_2016   & $0.867 \pm 0.001$ & $\textbf{0.873} \pm \textbf{0.001}$  & 0.938 $\pm$ 0.003 & \textbf{0.945} $\pm$ \textbf{0.003} \\
\bottomrule
\end{tabularx}
}
\end{sc}
% \end{small}
\end{center}
\caption{\small Comparison between MSE and Bradley-Terry losses on ProteinGym benchmark tasks using the CNN baseline model. Each row represents a data set and split combination. Numerical columns indicate the mean and standard deviation of test set metrics over 10 random initializations of the model. Asterisks indicate that unmodified portions of sequences were used in training data. Bold values indicate that a loss has significantly improved performance over all other tested losses ($p < 0.05$).}
 \label{tab:proteingym_results}
\end{table*}

% \newpage

% \section{Understanding the effectiveness of contrastive losses }
% Explain connection between Gaussian order statistics and gradients of contrastive losses.
\clearpage
\section{Derivation of fitness-epistasis uncertainty principle}\label{sec:derivation}

Here we show how to derive the fitness-epistasis uncertainty of Eq. \eqref{eq: uncertainty_principle}, starting from the uncertainty principle presented in Theorem 23 of \cite{Dembo1991}. When applied to the transformation $\textbf{f} = \bm{\Phi} \bm{\beta}$, this uncertainty principle is stated as:
\begin{equation}
    H(\textbf{f}) + H(\bm{\beta}) \geq \log\left(\dfrac{1}{M^2}\right)
\end{equation}
where $M = \max_{ij} |\Phi_{ij}|$. This can be further simplified by calculating the maximum absolute value of the elements in the Graph Fourier basis, $\bm{\Phi}$. As shown in \cite{Brookes2022}, this matrix for sequences of length $L$ and alphabet size $q$ is calculated as 
\begin{equation}\label{eq:construct_phi}
    \bm{\Phi} = \bigotimes_{i=1}^{L} \textbf{P}(q)
\end{equation}
where $\otimes$ denotes the Kronecker product, and $\textbf{P}(q)$ is an orthonormal set of eigenvectors of the Graph Laplacian of the complete graph of size $q$. Based on Eq. \eqref{eq:construct_phi}, we can make the simplification to the calculation of the maximum value in $\bm{\Phi}$ that 
\begin{equation} \label{eq:m_simplify}
    M = m^L
\end{equation}
where $m = \max_{ij} |P_{ij}(q)|$. Further, the value of $m$ can be determined for each setting of $q$ by considering the following calculation of $\textbf{P}(q)$ used in \cite{Brookes2022}:
\begin{equation}\label{eq:pq_def}
    \textbf{P}(q) = \textbf{I} - \dfrac{2\textbf{w}\textbf{w}^T}{||\textbf{w}||^2}
\end{equation}
where $\textbf{I}$ is the identity matrix and $\textbf{w} = \textbf{1} - \sqrt{q}\textbf{e}_1$, with $\textbf{1}$ representing a vector with all elements equal to one, and $\textbf{e}_1$ is the vector equal to one in its first element and zero in all other elements. The values of each element in \textbf{P}(q) can be straightforwardly calculated using Eq. \eqref{eq:pq_def}. In particular, we have:
\begin{equation}
    P_{ij}(q) = 
    \begin{cases}
    \frac{1}{\sqrt{q}}&\,\, \text{if} \,\, i=1\, \text{or}\, j=1 \\
    1 - \frac{1}{q - \sqrt{q}}& \,\, \text{if}\,\, i=j\neq 1 \\
    \frac{1}{q - \sqrt{q}} &\,\, \text{otherwise}.
    \end{cases}
\end{equation}
Now all that remains is determining which of these three values has the largest magnitude for each setting of $q$. For $q=2$,  only the first two values appear in the matrix, and both have magnitude $\frac{1}{\sqrt{q}}=\frac{1}{\sqrt{2}}$. For $q=3$, we can directly calculate that all three values are positive and we have $\frac{1}{q - \sqrt{q}} > \frac{1}{\sqrt{q}} > 1- \frac{1}{q - \sqrt{q}}$. For $q=4$, we can again directly calculate that all three values are equal to $ \frac{1}{\sqrt{q}} = \frac{1}{2}$. For $q > 4$, all three of the values are positive. Further, algebraic manipulations can be used to show that $q > 4$ implies $1 - \frac{1}{q - \sqrt{q}} > \frac{1}{q - \sqrt{q}}$ and $\frac{1}{\sqrt{q}} >\frac{1}{q - \sqrt{q}}$. Therefore, $1 - \frac{1}{q - \sqrt{q}}$ is the maximum value in $\textbf{P}(q)$ for all $q>4$.

Putting these results together with Eq. \eqref{eq:m_simplify}, the fitness-epistasis uncertainty principle simplifies to

\begin{equation}
    H(\textbf{f}) + H(\bm{\beta}) \geq L\log\left(\dfrac{1}{m^2}\right)
\end{equation}

where

\begin{equation*}
    m = 
        \begin{cases}
    \frac{1}{\sqrt{q}}&\,\, \text{if} \,\, q=2 \\
    \frac{1}{q - \sqrt{q}} &\,\, \text{if}\,\, q =3\\
    1 - \frac{1}{q - \sqrt{q}}& \,\, \text{if}\,\, q \geq 4 \\
    \end{cases}
\end{equation*}

which is the form presented in the main text.

\newpage
\end{document}